\documentclass[a4paper,11pt]{article}
\usepackage{pos}
\usepackage{subcaption}
\usepackage{siunitx}
\title{CMS track reconstruction performance during Run 2 and developments for Run 3}

\author*[\dag]{Walaa Elmetenawee}
\notes{\note{On behalf of the CMS Collaboration.}}
\affiliation{Università e INFN di Bari}
\emailAdd{walaa.elmetenawee@cern.ch}

\abstract{
An efficient and precise reconstruction of charged-particle tracks is crucial for the overall performance of the CMS experiment. During Run 2 of LHC, significant upgrades were made to the track reconstruction algorithms in order to accommodate for the high pileup environment and the installation of an upgraded pixel detector in 2017. This paper provides an overview of the iterative track reconstruction algorithm used in CMS during Run 2 and of the performance measured both with simulated and collision data. Developments are ongoing to further improve track reconstruction in Run 3, especially for what concerns the CMS high-level trigger, and the status of these improvements will be discussed.}

\FullConference{%
  40th International Conference on High Energy physics - ICHEP2020\\
  July 28 - August 6, 2020\\
  Prague, Czech Republic (virtual meeting)
}

\begin{document}
\maketitle

\section{Introduction}

The precise and efficient determination of charged-particle momenta is a critical component of the physics program of CMS~\cite{a}, as it impacts the ability to reconstruct the physics objects needed to understand pp collisions at the LHC. The central feature of the CMS apparatus is a superconducting solenoid of 6 m internal diameter, providing a magnetic field of 3.8 T. Within the solenoid volume there are the keys ingredients of the tracking system: a silicon pixel and a strip trackers. Muons are detected in gas-ionization chambers embedded in the steel flux-return yoke outside the solenoid. In March 2017, the original pixel detector was replaced with a new device, the “phase-1” pixel detector~\cite{e}, equipped with one additional barrel layer and one additional forward disk per side that features full 4-hit coverage in the tracking volume. This upgraded design places the innermost barrel layer even closer to the interaction point, 2.9 cm instead of 4.4 cm. The main motivation was to address the dynamic inefficiency driven by the internal readout chip buffers at high instantaneous luminosity and trigger rates. A comparison between the geometry of the old and new pixel detectors is shown in Figure 1.

\begin{figure}[ht]
\centering
\includegraphics[width=10cm]{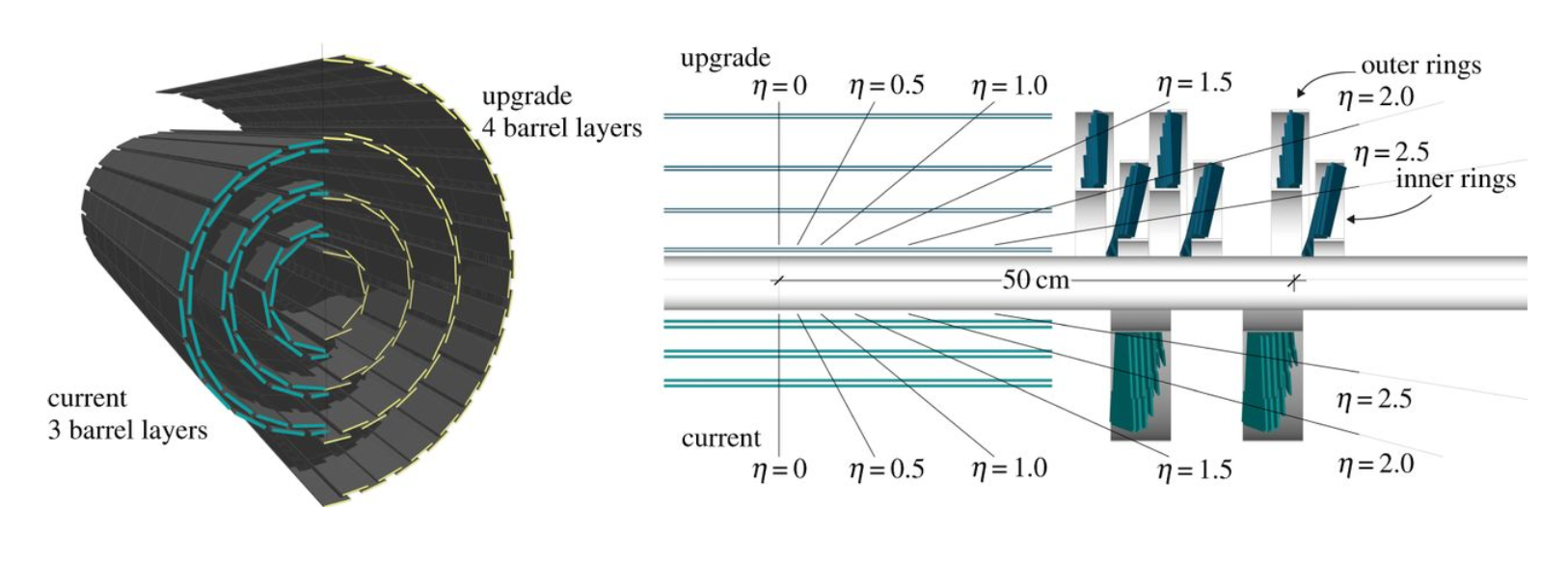}
\qquad
\caption{Left: Transverse-oblique view comparing the pixel barrel layers of the initial and upgraded CMS pixel detectors.  Right: Conceptual layout comparing the layers and disks in the two detectors.}
\end{figure} 

\section{Tracking performance in Run 2}
The CMS track reconstruction algorithm~\cite{b} ("Iterative Tracking"), is based on the combinatorial Kalman filter (CKF)~\cite{c}. The first set of iterations looks for tracks that are easier to find, like prompt track with relative high pT. Hits associated with reconstructed tracks are then removed from the set of hits. This procedure reduces the combinatorial complexity so that the more difficult kinematic regions tracks can be reconstructed. The track reconstruction can be described in a 4-step procedure: the first step is constructing seeds (also called "seeding"), which provides the initial estimate of the trajectory parameters. These proto-candidates are then propagated through the whole tracker, finding the compatible hit at each layer with the Kalman filter and updating the track candidate and its parameters ("building"). The tracks are then fitted after combining all associated hits ("fitting") and marked with quality flags ("selection"). The track reconstruction was updated in order to exploit the upgraded pixel detector features at the beginning of 2017. The expected tracking efficiency based on MC studies is shown in Figure 2, where events from  $t\bar{t}$ production with an average pileup of 35 have been used~\cite{d}. 

\begin{figure}[ht]
\centering
\includegraphics[width=6cm]{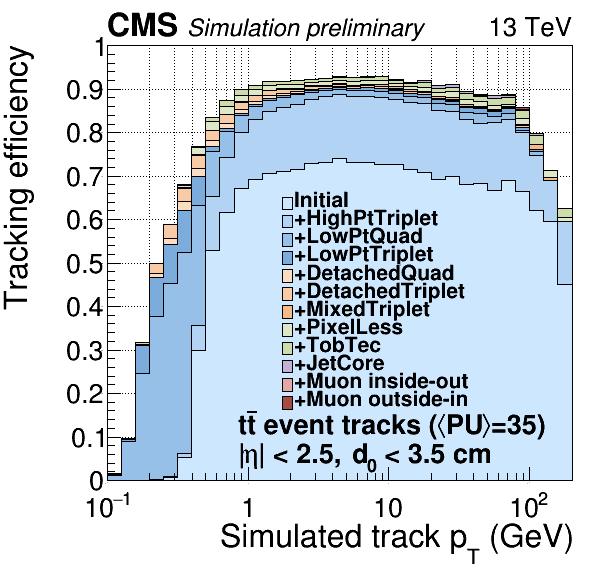}
\qquad
\caption{Tracking efficiency in MC as a function of track $p_{T}$}
\end{figure}

The most significant improvement introduced in the tracking comes from using the “Cellular Automaton” (CA) technique~\cite{n} to create track seeds using pixel-only hits, where hit pairs are formed between pixel detector layers. Hit triplets and quadruplets used for seeding are formed from compatible pairs after checking the seed compatibility with the beam-spot. A scheme of the algorithm is shown in Figure 3.

\begin{figure}[ht]
\centering
\includegraphics[width=10cm]{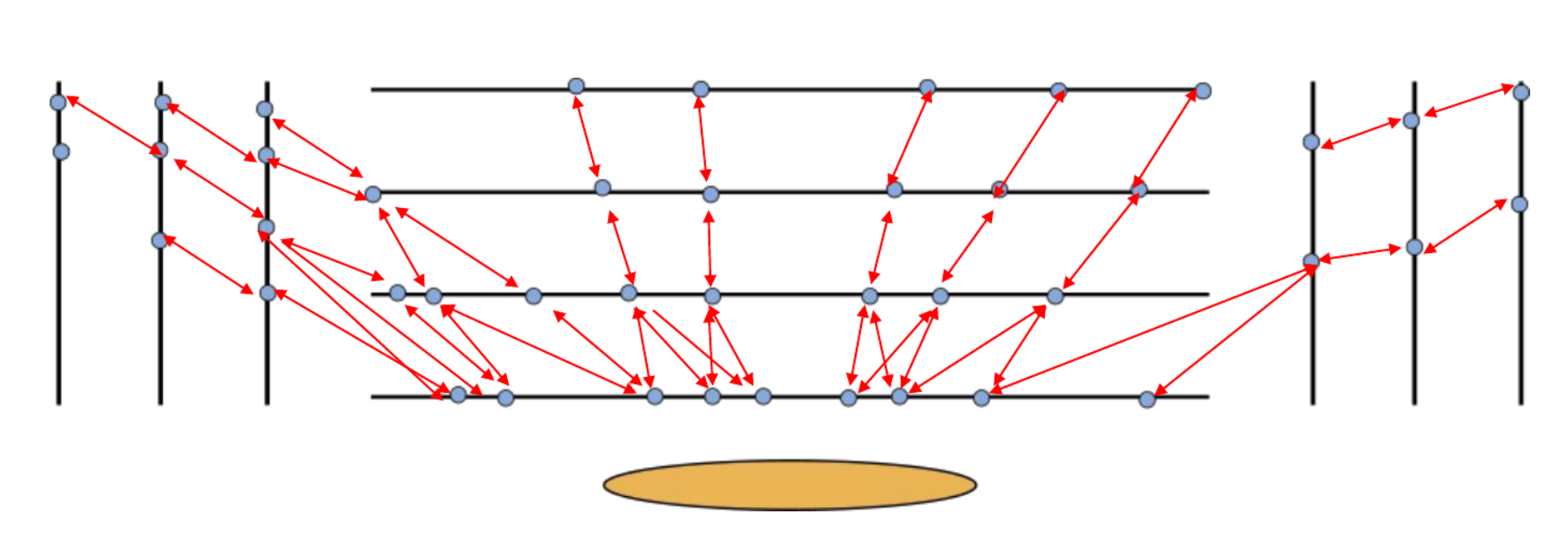}
\qquad
\caption{scheme of the Cellular Automaton track seed algorithm.}
\end{figure} 

A comparison between the track reconstruction used in 2016 (with the phase-0 pixel detector) and 2017 (with the phase-1 detector) is shown in Figure 4. A considerable increase in tracking efficiency is observed, mainly in the forward regions of the detector. Meanwhile, the track fake rate is considerably reduced. With additional pixel layers, a significant slowing down in the seed finding step would be expected. Using the CA for track seeding instead improves the timing. The comparison of these scenarios is shown in Figure 5 (left). The reduction in the fake rate already at seeding level leads to reduced time spent on the pattern recognition, independent of the seed algorithm, as shown in Figure 5 (right).

\begin{figure}[ht]
\includegraphics[width=6cm]{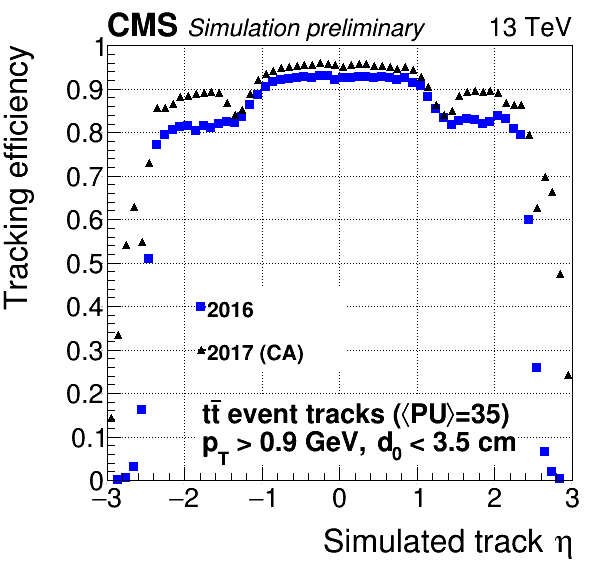}
\qquad
\includegraphics[width=6cm]{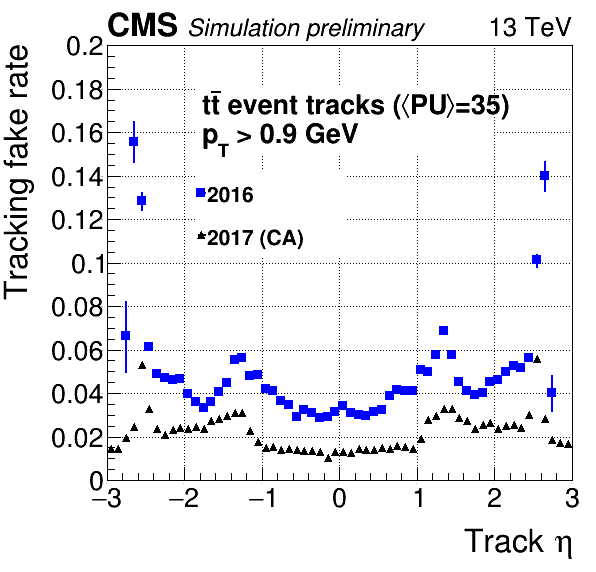}
\caption{Track reconstruction performance as a function of simulated track $\eta$ for 2016 and 2017 detectors, latter with Cellular Automaton (CA) seeding. The tracking efficiency (left) and fake rate (right)~\cite{d}.}
\end{figure}

\begin{figure}[ht]
\includegraphics[width=6cm]{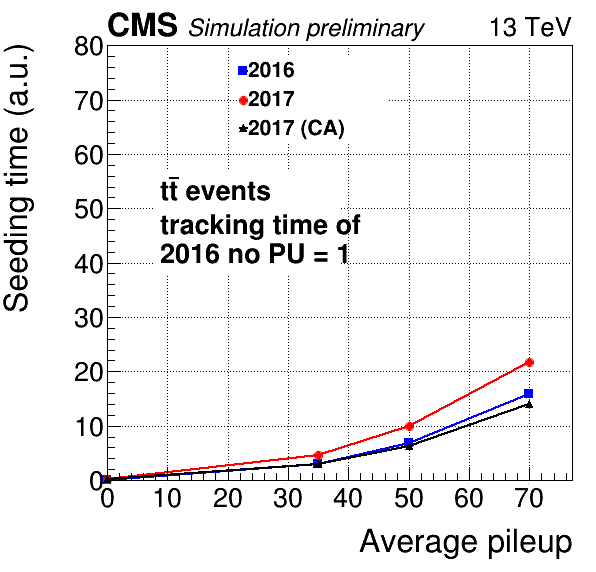}
\qquad
\includegraphics[width=6cm]{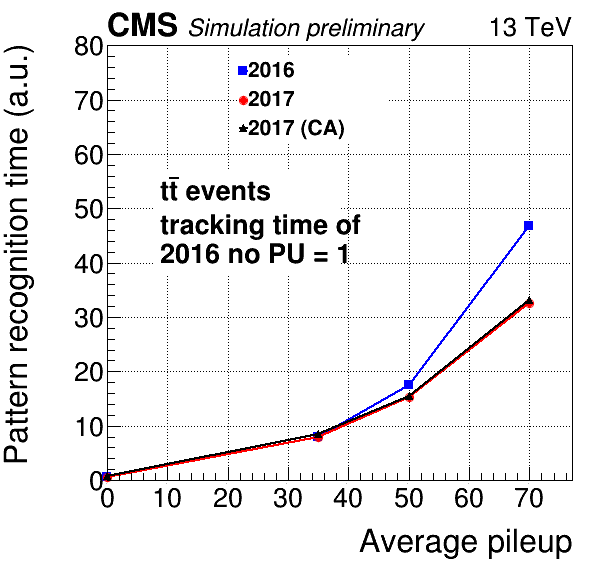}
\caption{Seeding (left) and pattern recognition (right) time in simulation as a function of average pileup for both the 2016 and 2017 detector~\cite{d}.}
\end{figure}

The tracking performance is measured in data applying the tag-and-probe technique ~\cite{j}. The efficiency is estimated using events with a Z boson decays into a pair of muons. The so-called "tag" muon is reconstructed using both the muon chambers and the tracker and identified by stringent requirements. The other muon ("probe"), instead, is reconstructed using only hits from the muon system and identified by looser requirements. Probes are then sorted into two categories, passing and failing, considering whether or not they can be matched to at least one track in a cone ($\Delta R$ < 0.3) around the probe muon. The efficiency is calculated as the ratio between the passing probes and the total number of probes. The tracking efficiency of about 99.9 $\%$ for tracks associated with muons has been guaranteed in the whole muon pseudorapidity acceptance through the whole Run2 data taking, as shown in Figure 6 (left). There is a very mild dependence on the pileup conditions, as highlighted in the distribution in Figure 6, which is mainly driven by a pixel dynamic inefficiency in the readout chip of the modules in the innermost barrel layer.

\begin{figure}[ht]
\includegraphics[width=6cm]{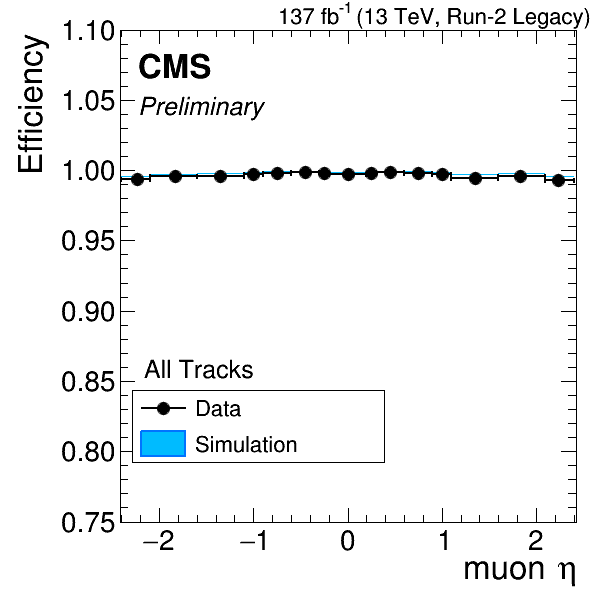}
\qquad
\includegraphics[width=6cm]{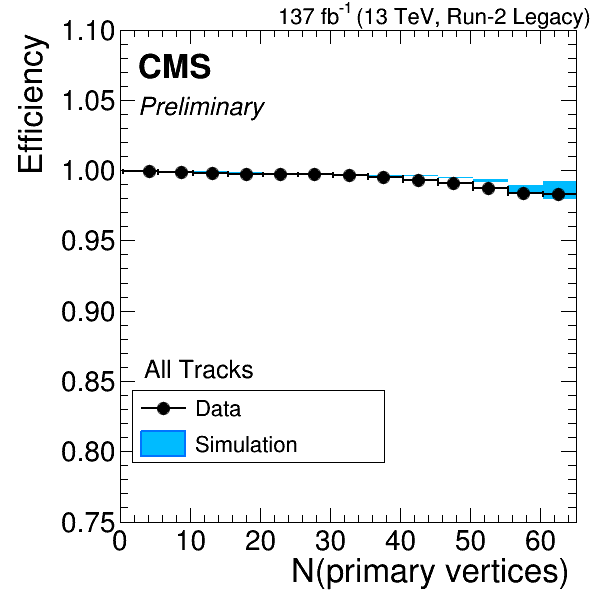}
\caption{The tracking efficiency in Run 2 Legacy as a function of pseudo-rapidity $\eta$ (left) and the number of primary vertices (right) using the tag and probe technique.~\cite{f}.}
\end{figure}

At high momentum, the impact parameter resolution is dominated by the position resolution of the innermost hit in the pixel detector, while the resolution is degraded by multiple scattering at lower momentum. The transverse track impact parameter resolution measured with respect to the beam-spot position~\cite{i} is shown in Figure 7 as a function of track $p_{T}$ (left) and track $\eta$ (right). As expected, the closer position of the innermost layer to the beam pipe with respect to the phase-0 detector~\cite{b} lead to a notable improvement to the IP resolution, and in addition, the new layout allowed a better coverage in the pseudorapidity.

\begin{figure}[ht]
\includegraphics[width=6cm]{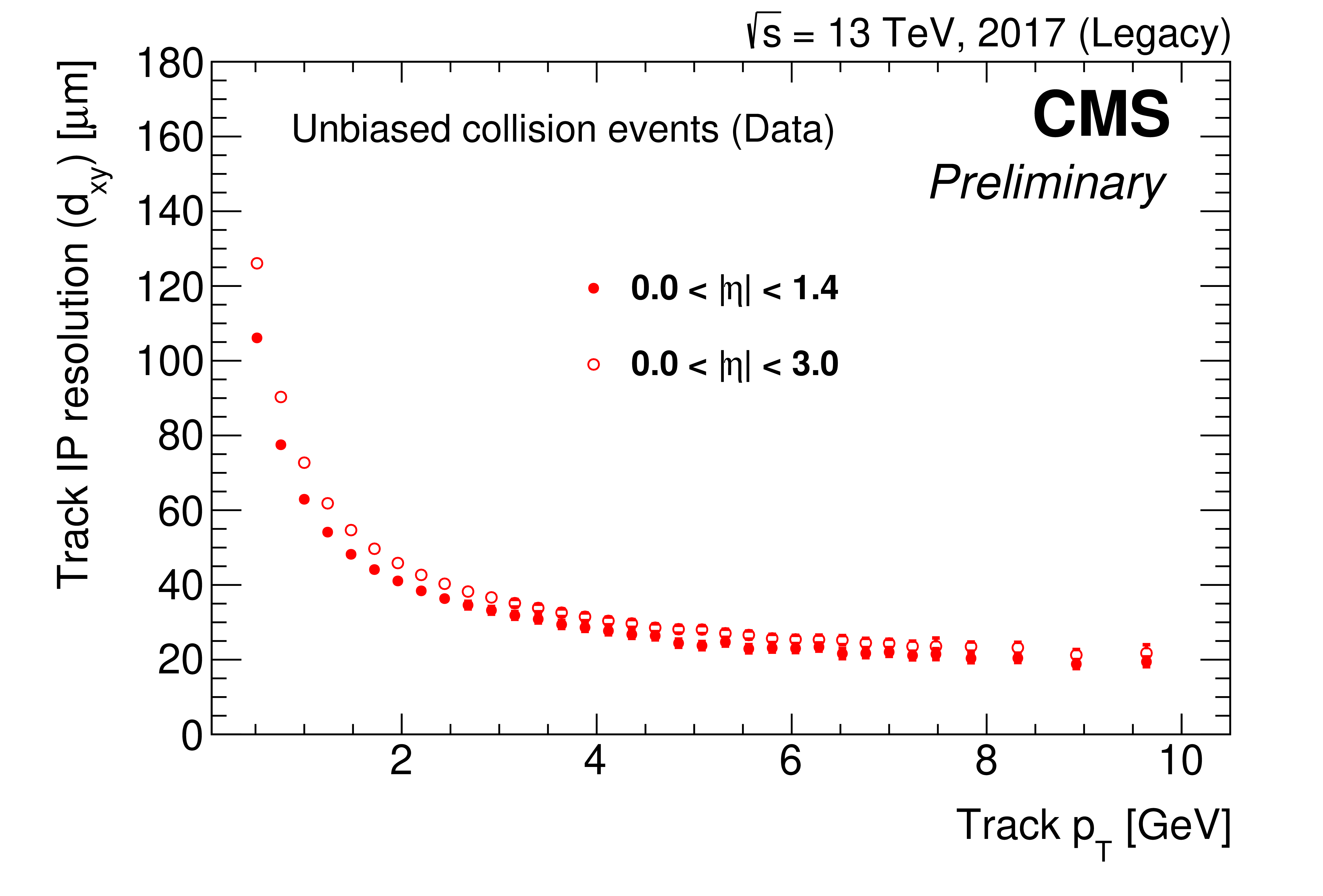}
\qquad
\includegraphics[width=6cm]{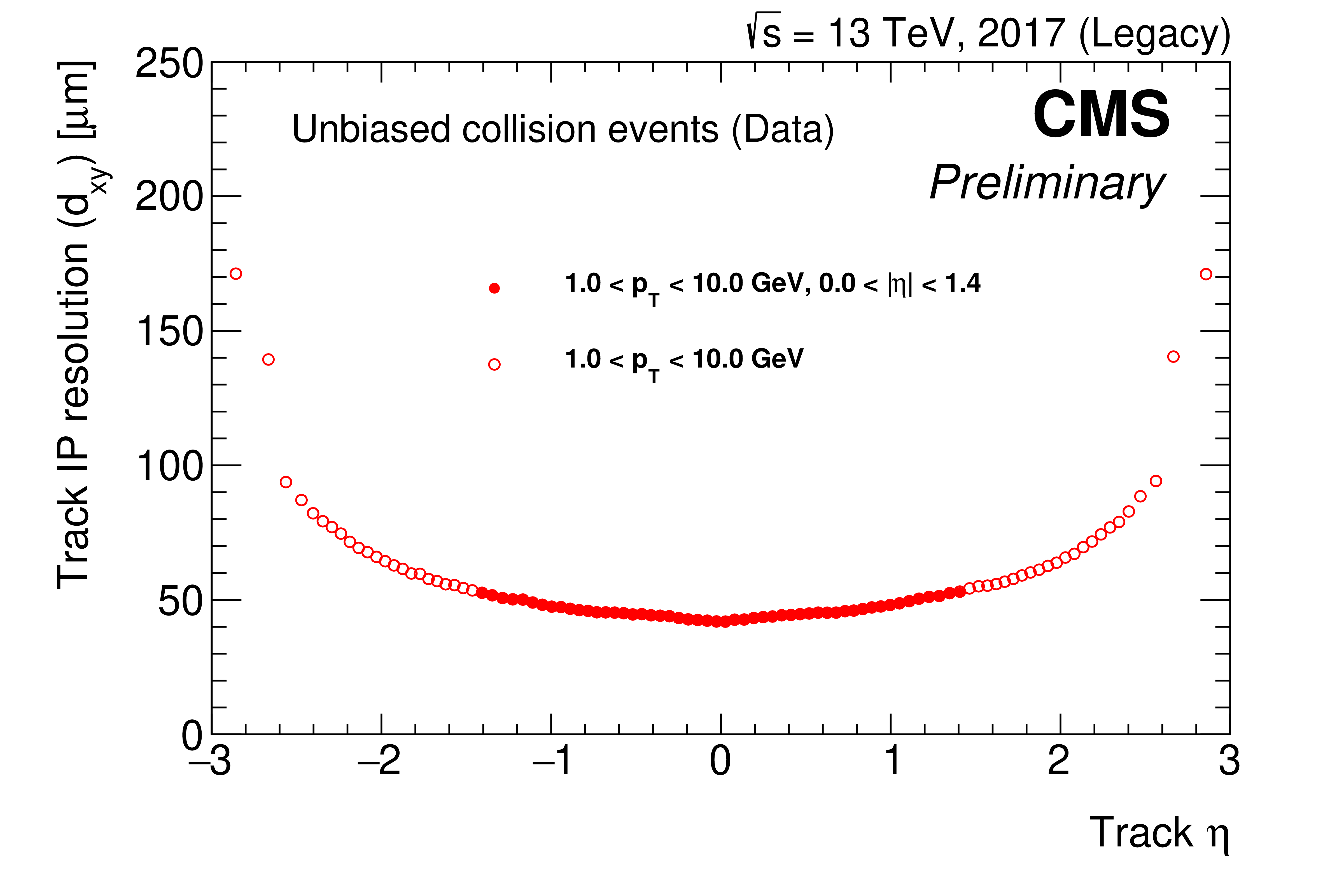}
\caption{Transverse impact parameter (IP) resolution measured as a function of track $p_{T}$ (left) and track $\eta$ (right). The region of $\lvert\eta\lvert$ < 1.4 corresponds to the coverage of the Phase-0 barrel Pixel detector~\cite{g}.}
\end{figure}

\section{Tracking at High-level trigger (HLT)}
The HLT gathers information from all the CMS detectors, and among them, the tracker detector has a crucial role in two of the first steps of the reconstruction used for the event selection: the track reconstruction and the primary vertex finding. Tracking at HLT is performed iteratively and uses a track reconstruction software that is almost identical to the one used in offline reconstruction~\cite{b}, but it has to fulfill stringent CPU timing constraints. The general track reconstruction in the HLT consists of three iterations. The first two require a maximum of four consecutive hits in the pixel detector to seed the tracking. These iterations target first high and then low $p_{T}$ tracks and exploit the whole volume of the detector. The third iteration relaxes the number of hits in the track seeds to three; it is restricted to jet candidates vicinity identified from calorimeter information and the tracks reconstructed in the two previous iterations~\cite{k}.

The contributions to the total efficiency from the different tracking iterations in 2018 are shown in Figure 8. Compared to the performance with a perfect detector, the efficiency loss is concentrated in the region around $\phi$= 0.6, where in 2018 data taking there were a concentration of pixel inactive modules in two adjacent layers. The doublet-seeded iteration was introduced in order to recover the efficiency in these kind of pathological regions. The efficiency is robust against pileup, decreasing slightly with increasing the number of interactions.

\begin{figure}[ht]
\includegraphics[width=6cm]{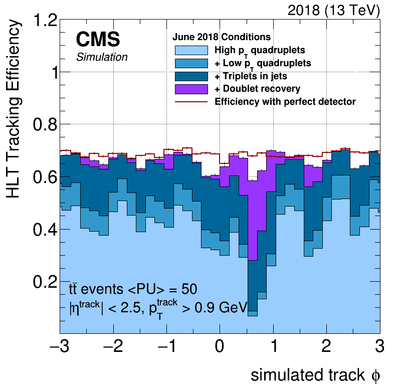}
\qquad
\includegraphics[width=6cm]{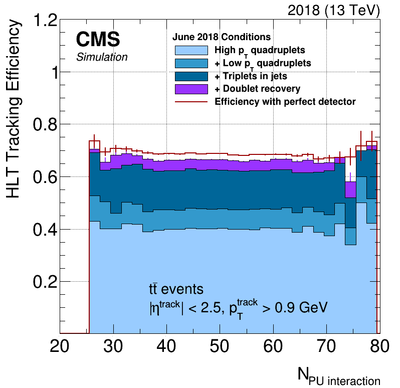}
\caption{Tracking efficiency measured with respect to Monte Carlo truth information as a function of simulated track $\phi$ (left) and the number of PU interactions (right).~\cite{k}.}
\end{figure}

\section{Developments targeting Run 3}
Increasing the luminosity and, consequently, the accumulated radiation will pose new challenges for the CMS detector in the upcoming years and, specifically, for track reconstruction in the pixel detector. Several developments are ongoing targeting Run 3; improvements in the mitigation strategy, the usage of deep neural networks (DNN) for the track selection, a seed cleaning with Convolutional Neural Network (CNN)~\cite{l}, and the usage of DeepCore track seeding in dense environments with CNN~\cite{h}. In this section, some of these developments will be briefly discussed.

During track seeding, compatible pairs of hits from different pixel detector layers are subsequently fed to higher-level pattern recognition steps. However, the set of compatible hit pairs is heavily affected by the combinatorial background resulting in the tracking algorithm following steps to process a significant fraction of fake doublets. A possible way to reduce this effect is by considering that each hit is a cluster of pixels with its own shape to check the compatibility between two hits. Therefore the task of fake rejection can be seen as an image classification problem for which CNNs have been widely proven to provide reliable results. The preliminary tests confirmed that using the CNN in the track seeding would practically leave unchanged the downstream track reconstruction performances while heavily reducing the combinatorial fakes and consequently improving the timing performance~\cite{l}, which will be ideal for HLT tracking.

The tracking inside the jet core becomes inefficient for high transverse momentum jets due to the collimated environment that produces merged clusters from different tracks on the pixel detector. The splitting of the merged cluster is inefficient, and it degrades the quality of the seeding of the CMS tracking algorithm, based on a combinatorial Kalman Filter. Using the Convolutional Neural Network (CNN) allows to skip the pixel clustering and produces the track-seeds directly from the four layers raw pixel information, as shown in Figure 9~\cite{h}. The improvement given by DeepCore to the track reconstruction is shown in Figure 10, where the performance with the standard jetCore algorithm and the one with DeepCore are compared. Using the DeepCore algorithm almost cancels the seeding inefficiencies and shows better performance than the standard seeding algorithm in such a dense environment. 
\begin{figure}[ht]
\begin{subfigure}{.19\textwidth}
  \centering
  \includegraphics[width=.8\linewidth]{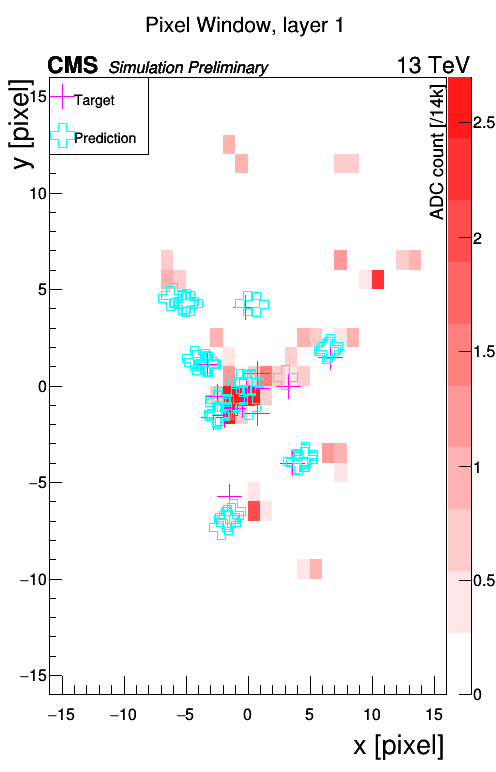}
  \caption{}\label{fig:sfig1}
\end{subfigure}%
\begin{subfigure}{.19\textwidth}
  \centering
  \includegraphics[width=.8\linewidth]{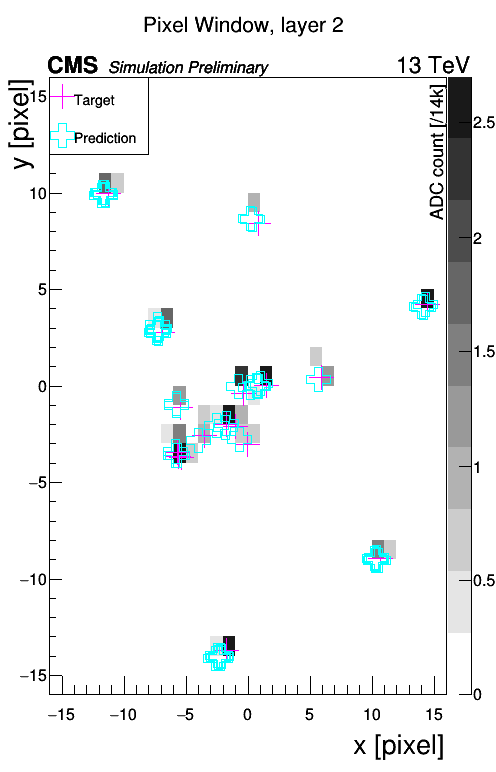}
  \caption{}\label{fig:sfig2}
\end{subfigure}
\begin{subfigure}{.19\textwidth}
  \centering
  \includegraphics[width=.8\linewidth]{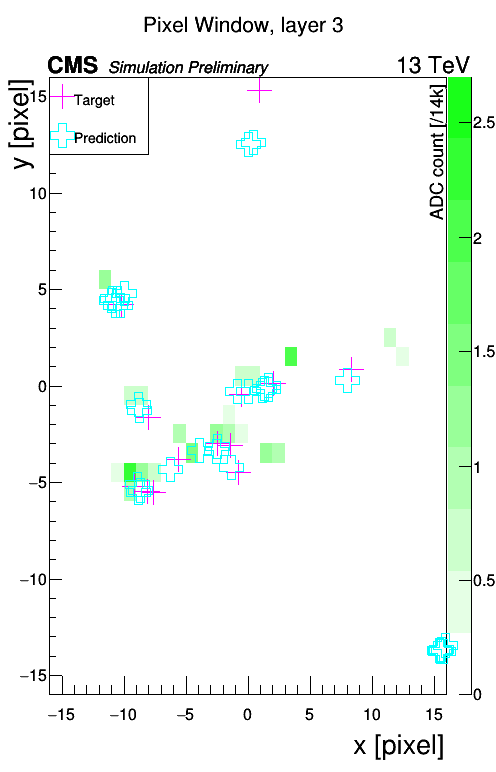}
  \caption{}\label{fig:sfig3}
  \end{subfigure}
  \begin{subfigure}{.19\textwidth}
  \centering
  \includegraphics[width=.8\linewidth]{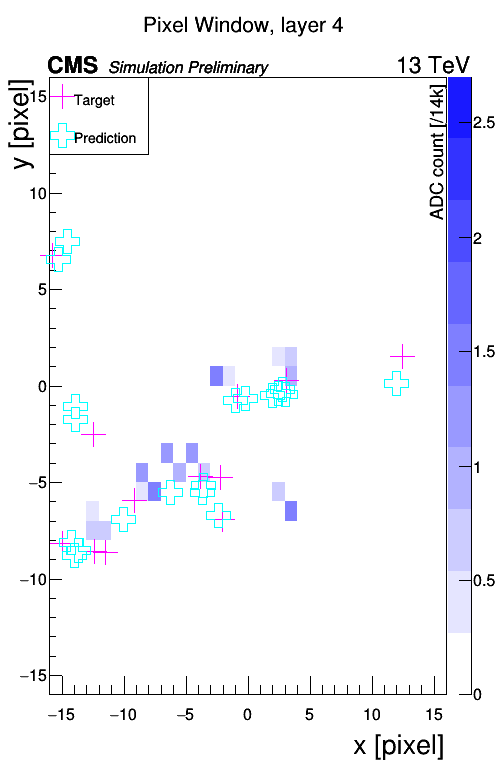}
  \caption{}\label{fig:sfig4}
  \end{subfigure}
  \begin{subfigure}{.19\textwidth}
  \centering
  \includegraphics[width=.8\linewidth]{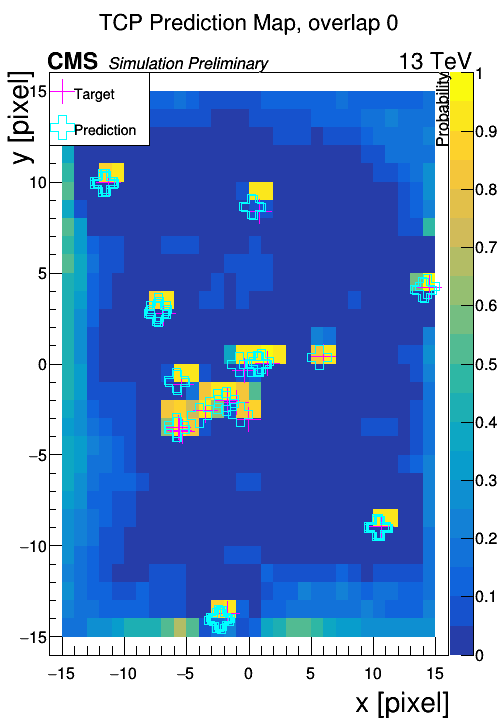}
  \caption{}\label{fig:sfig5}
\end{subfigure}
\caption{The pixel maps of the four detector layers (a-d) are used as input for the neural network. The pixel map (e) is the map of the predicted crossing point on the window of layer 2, expressed as a probability, with the predictions and targets' crosses.}
\label{fig:fig}
\end{figure}
\begin{figure}[ht]
\centering
\includegraphics[width=6cm]{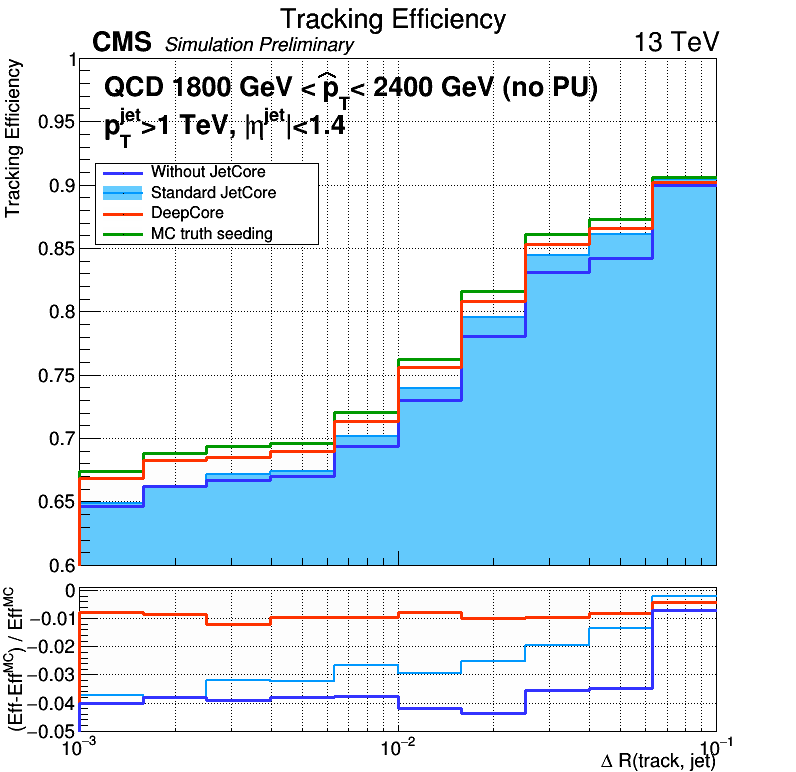}
\qquad
\caption{The tracking efficiency in the jet core region ($\Delta R$ < 0.1, between the reconstructed jet axis and the simulated track direction). The lower pad shows the differences between various tracking efficiencies and the MC truth seeding one, divided by the MC truth seeding efficiency.}
\end{figure} 
\section{Conclusions}
Despite the challenging environment at the LHC in Run 2, the CMS tracker features robust performance and efficient tracking. Upgrades to the track seeding algorithms allowed to exploit of the upgraded pixel detector within the same CPU time budget, resulting in higher efficiency and lower fake rate. 

The Run 3 environment is expected to be more extreme; thus, further improvements in the mitigation strategy, track seeding, seed cleaning, track selection and dedicated tracking step for displaced signatures are under development.

\end{document}